\documentstyle[12pt]{article}
\begin{document}

\vbox{\begin{flushright}
CINVESTAV-FIS-50-96
\end{flushright}

\title{ Tests of Higgs and Top Effective Interactions }
\author{ J. L. D\'\i az-Cruz $^a$, M. A. P\'erez$^b$ and
J. J. Toscano$^c$  \\
 (a) Instituto de F\'\i sica, BUAP,  \\
 Apdo. Postal J-48, 72500 Puebla, Pue., Mexico \\
 (b) Departamento de F\'\i sica, CINVESTAV-IPN \\
 Apdo. Postal 14-740, M\'exico, D.F. 07000.    \\
  (c) Facultad de Ciencias F\'\i sico-Matem\'aticas, BUAP \\
      Apdo. Postal 1152, 72000 Puebla, Pue., M\'exico }
\maketitle
\begin{abstract}
We study the possibility to detect heavy physics effects 
in the interactions of Higgs bosons and the top quark at future
colliders using the effective Lagrangian approach, 
where the SM Lagrangian is modified
by non-renormalizable operators that are invariant under 
the full strong and electroweak group. The modification  
of the interactions may enhance the 
production of Higgs bosons at hadron colliders 
through the mechanisms of gluon fusion and associated 
production with a W boson or $t\bar{t}$ pairs. 
The most promising signature is through the decay of the
Higgs boson into two photons, whose branching ratio is also
enhanced in this approach. As a consequence of our analysis we get
a bound on the chromomagnetic
dipole moment of the top quark.

\end{abstract}

}

\newpage

\normalsize


{\bf 1. Introduction.}
The standard model (SM) \cite{smori} has been tested with 
success at the level of radiative corrections at LEP \cite{lepa}.
However, several properties of the elementary particles 
are not explained within the model, like
the large value found for the top quark mass or
the nature of the Higgs mechanism, responsible 
for the generation of masses. 
Because of the lack of any experimental evidence on 
the Higgs boson, almost anything could be said
about its nature. For instance, that it could be the remnant
of some new physics that governs the world 
at deeper distances. 

The present generation of colliders (LEP, FNAL)
has tested only some portion of the parameter space of 
the Higgs sector within the SM and beyond;
for the SM it is found that $m_H>67$ GeV \cite{lepa}.
It is expected that the
next generation of colliders (LHC, NLC)
will be decisive to further test the SM
and to discover the nature of the
Higgs mechanism.

The framework of effective Lagrangians, as a mean to
parametrize physics beyond the SM in a model independent
manner, has been used extensively recently \cite{effecl}.
Two main cases have been discussed in the literature,
the decoupling case \cite{effeca}, which assumes the existence
of a light Higgs boson, and the non-decoupling
case \cite{effecb}, where no Higgs boson is included at all. 
We shall consider here the decoupling case, 
which considers the SM as the low-energy 
limit of a weakly coupled and renormalizable full theory.
Within this approach, the effective Lagrangian is constructed
by assuming that the virtual effects of new physics
modify the SM interactions in such a way that they are 
parametrized by a series
of higher-dimensional nonrenormalizable operators
written in terms of the SM fields. These operators
respect the SM symmetries and are suppressed by
inverse powers of the high-energy 
new physics scale $\Lambda $ \cite{georg}.

In this paper we study how the detection of the Higgs bosons 
at hadron colliders is affected when its couplings with 
gauge bosons and the top quark are modified 
within the effective Lagrangian approach. 
We shall focus mainly in the so called intermediate
mass-region $(m_Z<m_H<2 m_Z)$, which is the prefered region by
the analysis of SM radiative corrections 
\cite{jellis}. We find that the 
modification to the SM interactions may enhance 
the cross-sections for the production of Higgs bosons 
through the reactions of gluon fusion and the 
associated production with a W boson or $t\bar{t}$ pairs. 
Using a perturbative criteria and our result
on the gluon fusion mechanism, we obtain a bound on the
chronomagnetic dipole moment of the top quark.
\vskip1cm

{\bf 2. The effective Lagrangian.}
The effective Lagrangian for Higgs $(H^0)$ and top $(t)$
interactions with gauge bosons ($W,Z, \gamma$) and
gluons ($g$), can be expanded as follows:
\begin{equation}
{\cal L}_{eff}=  {\cal L}_{SM} 
+ \sum {\frac{\alpha_i}{\Lambda^n} } O^i_n,
\end{equation}

\noindent where ${\cal L}_{SM}$
denotes the SM renormalizable Lagrangian.
The terms $O^i_n$ are higher-dimension
(non-renormalizable) $SU(3) \times SU(2)_L \times U(1)$
invariant operators and $\alpha_i$ are unknown parameters,
whose order of magnitude can be estimated because
gauge invariance makes possible to establish  the
order of perturbation theory in which each operator
can be generated in the full theory \cite{effecc}. 
The coefficients of the operators which are
generated at tree-level will be suppressed only by products of
coupling constants, whereas those that can be generated
at the one-loop level, or higher, will be further suppressed by a
typical $16\pi^2$ loop factor. 

One consequence of assuming that the full gauge symmetries of
the SM should be respected by the new operators
is that the lowest-dimension operators in (1) are of dimension-6.  
As we shall discuss below, in the case of the top quark
this implies that the value expected for the
chromomagnetic dipole moment may lay beyond the sensitivity of
future experiments, and thus the bounds obtained in the 
literature for this moment
\cite{topeff} may be suppressed by an additional
factor $v/\Lambda$, with $v=254$ GeV 
the electroweak scale. This situation is similar to that
found in the study of the 3-point vertices within the context
of the effective Lagrangian approach: the use of the full
SM gauge symmetry imposes a stronger bound 
\cite{tptgv} than the one obtained with the mere use of 
$U(1)_{em}$ gauge invariance \cite{ginve}.

In this paper we are interested in the 
gluon fusion mechanism $(gg\to H^0)$, which in the SM occurs  
at the one-loop level (Fig. 1).
This interaction involves 
both the strong and the Yukawa sectors of the SM.
The top quark gives the leading contribution 
to the loop in the SM. The virtual effects of new physics could modify 
the $ggH$ interaction, and if the new physics
is described at the scale $\Lambda$ by
a perturbatively renormalizable theory, then its
effects will decouple in the limit $v/\Lambda \to 0$.
There are three possible sources of change for the $ggH$ effective vertex:
the $H^0t\bar{t}$ interactions (Fig.. 1-ii),  
the QCD vertex $gt\bar{t}$ (Figs. 1-iii,iv,v), 
and the effective 
contact term for the vertex $ggH$ induced by 
new physics effects (Fig. 1-vi).

We denote by $t_{L,R}$  the chiral components of the
top quark; q corresponds to the top-bottom doublet, 
$\phi, \, G$ are used for
the Higgs doublet and gluon fields, and $W,B$ for the
electroweak gauge bosons. We shall employ the notation
$O^{n,i}_X$ for an operator of type X, with dimension n, and
$i=t,l$ for tree- and loop-induced operators, respectively.
The dimension-6 operator

\begin{eqnarray}
 O^{6,t}_{t\phi}=({\phi}^{\dagger} \phi) \bar{q}t_R \tilde{\phi} 
\end{eqnarray}
modifies the Yukawa interaction of the top-Higgs system .
One operator that leads to modifications
of the $gt\bar{t}$ interaction, induced 
at the one-loop level, is given by:
\begin{equation}                                     
 O^{6,l}_{\phi Gt}=i \bar{q} G^{\mu\nu}
                    \sigma_{\mu\nu} \tilde{\phi} t_R,  
\end{equation}
where $G^{\mu\nu}=\lambda^a G^{a\mu\nu} $, with $\lambda^a$
denoting the Gell-Mann matrices. Also
$\sigma_{\mu\nu}=(i/2)[\gamma_{\mu},\gamma_{\nu}]$, with  
$\gamma_{\mu}$ denoting the Dirac matrices.\footnote{We
have included all the relevant operators which induce the
$H\bar{t}t$, $Hgg$, $HWW$, and $HZZ$ vertices. On the 
other hand, there are also two other operators that can change 
the $gt\bar{t}$ vertex:
$ O^{6,l}_{qG}=i \bar{q} G^{\mu\nu}\gamma_{\mu}D_{\nu}q$, 
and $O^{6,l}_{tG}=i \bar{t}_R G^{\mu\nu}\gamma_{\mu}D_{\nu}t_R $,
however, in order to keep the analysis as simple as possible
we shall not consider them.}

A contact term for the effective vertex $ggH$  
is induced also by operators of dimensions 6 and 8,
generated at one-loop and tree levels, respectively,
and have the following structure:
\begin{eqnarray}
 O^{6,l}_{\phi G}&=&{\phi}^{\dagger} \phi G^{a\mu\nu} G^a_{\mu\nu}  \\
 O^{8,t}_{\phi G}&=&({\phi}^{\dagger} \phi)^2 G^{a\mu\nu} G^a_{\mu\nu} 
\end{eqnarray}

The operators (2-3) may also lead to some changes in the 
cross-section for the associated production of 
the Higgs boson with $t\bar{t}$ pairs, which is also
a relevant mechanism in the intermediate mass-region.

Another interesting aspect of the Higgs phenomenology,
namely the associated production of the Higgs with a W boson,
could be modified by the following set of 
dimension-6 operators, induced at tree-level, 
which modify the interaction of the Higgs boson with
the $W,Z$ bosons,
\begin{eqnarray}
 O^{6,t}_{\phi 1}={\phi}^{\dagger} \phi 
 (D_{\mu}\phi)^{\dagger} D^{\mu}\phi,  \\
 O^{6,t}_{\phi 3}=({\phi}^{\dagger} D_{\mu}\phi) 
 (D^{\mu}\phi)^{\dagger} \phi  ,
\end{eqnarray}
with the covariant derivative given, in general, by:
\begin{equation}                                     
D_{\mu}=\partial_\mu- {i\over 2} (g'YB_\mu+g\tau^i W^i_\mu 
+g_s\lambda^a G^i_\mu).
\end{equation}
The field tensors for the gluons and electroweak gauge
bosons are written as:
$G^a_{\mu\nu} = \partial_\mu G^a_\nu - \partial_\nu G^a_\mu
         +g_sf^{abc}G^b_{\mu}G^c_{\nu} $,
$W^i_{\mu\nu} = \partial_\mu W^i_\nu - \partial_\nu W^i_\mu +
g\varepsilon_{ijk} W^j_\mu W^k_\nu,$
$B_{\mu\nu} = \partial_\mu B_\nu - \partial_\nu B_\mu $. These 
operators modify the SM Feynman rules, which in turn will
affect the decay rates and production cross-sections of the
Higgs boson. The derivation of these rules, in terms of
physical fields, is quite lenghty and will not be reproduced
here.
In what follows, we shall write directly our results for
the formulae of the relevant reactions.

{\bf 3. Testing Higgs and top interactions.}
One of the aims of the present paper 
is to study the operators that modify 
the production of Higgs bosons through gluon fusion,
which is a relevant production mechanism 
in the so-called intermediate $(m_Z<m_H<2 m_Z)$ and heavy mass 
regions $(2 m_Z < m_H < 600 \ GeV)$ \cite{hixhunt}.  
The graphs that contribute to the
effective vertex $ggH$  are shown in Fig. 1, where
the dots denote the new vertices induced by the operators
of the effective Lagrangian.

Taking into account the operators (2-7), 
we have computed the decay width 
$\Gamma(H \to gg)$ in the effective Lagrangian approach, 
and the result is compared with the SM value 
through the following ratio: 
\begin{equation}
R_{Hgg}=\frac{\Gamma_{eff}(H \to gg)}{\Gamma_{SM}(H \to gg)}
       = \frac{|F_{eff}|^2}{|F_{SM}|^2},
\end{equation}
where
\begin{eqnarray}
F_{eff}&=&[1-\frac{3}{2\sqrt{2}}z_1^2z_2\alpha^{6,t}_{t\phi}] F_{SM}
 +[\frac{gm_tz_1^2}{4\sqrt{2} \pi^2g_s m_W}\alpha^{6,l}_{\phi tG}] 
                       F_{\phi tG}  \nonumber      \\
       & & -2z_1^2[\alpha^{6,l}_{\phi G}+
         ( \frac{4\pi v}{\Lambda})^2 \alpha^{8,t}_{\phi G} ],
\end{eqnarray}
with $z_1=v/\Lambda, \, z_2=v/m_t$, and $g,g_s$ denote 
the weak and strong coupling constants and
\begin{eqnarray}
F_{SM}&=&-2t[1+(1-t)I(t)],     \\
F_{\phi tG} &=&-5\log \frac{\Lambda^2}{m^2_t}+11-6t-2tI^2(t)  \nonumber \\
          & & - 2(3t-5)\sqrt{t-1}I(t),
\end{eqnarray}
with $t=\frac{4m^2_t}{m^2_H}$ and
the function $I(t)$ is given by
\begin{eqnarray}
I(t)&=&\arctan (1/\sqrt{t-1}),    \mbox{ for $t >1$ }, \nonumber  \\
    &= & [log (\eta_{+} / \eta_{-})+i\pi]/4, \mbox{ for $t <1$ }, 
\end{eqnarray}
with $\eta_{\pm}=1 \pm \sqrt{1-t}$. The quantity $R_{Hgg}$ is useful 
in order to evaluate the cross-section for the gluon 
fusion reaction. At any energy it is given in terms
of the SM result by $\sigma_{eff}= R_{Hgg} \sigma_{sm}$, 
for a given Higgs mass.

It seems convenient to comment here on a technical point about the
calculation, namely the fact that the contribution from the
operator $O^{6,t}_{t\phi }$ to the loop (graph 1-ii) is finite. 
This happens because,
after spontaneous symmetry breaking,
the effective interaction $H^0t\bar{t}$ is of dimension 4.
On the other hand, the operator  $O^{6,l}_{\phi tG }$ induces 
a divergent contribution through the dimension-5 vertex 
$gt\bar{t}$. This divergence can be absorbed through 
the modern criteria of renormalizability \cite{georg,newren} 
for effective Lagrangians.\footnote{The effective Lagrangian is constrained by
Lorentz and gauge invariance. These symmetry principles constraint in the
same way the ultraviolet divergences of the theory. Since the 
effective Lagrangian already include the infinite tower of interactions 
allowed by these symmetries, then the counterterms needed to cancel every 
ultraviolet divergence are already included in the theory. On the other
hand, predictibility of the theory depend strongly on the energies used
($E \ < \ \Lambda$) and the experimental accuracy.}

In our calculation, we have used a $\bar{MS}$-like scheme,
where the scale parameter $\mu$, associated to the
dimensional regularization procedure, 
is set equal to the energy scale $\Lambda$.

In order to proceed further,
we need to know the values of the coefficients 
$\alpha_i$. Because QCD (as well as QED) has an exact 
gauge symmetry, it seems reasonable to assume that 
each time the gauge fields appears in the operators $O_X$, 
one should put the gauge coupling constants as a factor in  
$\alpha_X$.
Thus, for a tree-level induced operator that contains the 
gauge fields n-times, one can write:  $\alpha_X=g^n_s\eta_X$,
whereas for the case of one-loop induced operators
it will take the form $\alpha_X= g^n_s\eta_X /16\pi^2$. 
The factor $\eta_X$ is left free to include the products of other
coupling constants (of broken symmetries or Yukawa coupings),
as well as possible group factors for the heavy fields that could
lay in larger representations of the QCD group. 
However, in order to present our numerical results, we shall fix $\eta=\pm1$
and will choose the appropriate sign combinations 
that give the maximum and minimum values for
the quantities of interest.

A priori, one could expect that the tree-level dimension-six
operator $O^{6,t}_{t\phi }$ should give the largest contribution 
to the loop,  
while the contribution of the operator $O^{6,l}_{\phi Gt}$
should be suppressed because it is one-loop generated. 
The loop-induced operator $ O^{6,l}_{\phi G}$ 
is expected to be suppressed too.
However, for small values of the scale 
$\Lambda$, the dominant contribution may arise from 
the contact terms, as we shall explain bellow. 

The results are presented in Table 1, where we display the 
effects of each operator on the ratio (9). We can appreciate that
there are significant changes coming from all the operators 
discussed here. The largest effect
is due to the 8-dimensional operator $O^{8,t}_{\phi G }$, 
which may enhance the ratio by a factor about 5.3,
although this effect is only valid
for $\Lambda \leq 3$ TeV. When both $\Lambda$ and $m_H$ grow, the 
largest effect comes from the operator $O^{6,t}_{t\phi }$,
as it was expected from the above rules.

The corresponding increase in the cross-section seems so large, 
that it might be possible to look for it at Tevatron. 
We found that the cross-section is only
$\sigma\simeq 5.3$ pb, for $\sqrt{s}=1.8$ TeV, $m_H=100$ GeV. Thus,
with a yearly integrated luminosity of $100$ $pb^{-1}$, we should expect
about 530 Higgs bosons per year. If the decay into a photon pair
receives a similar enhancement in the effective Lagrangian approach
($B.R. (H\to \gamma\gamma)\simeq 5 \times 10^{-3}$, as it was found in
\cite{toscaa}), then there will be only about 2.5 events per year, which
seems difficult to search for. However, if the luminosity is increased, as
it is expected in the upgraded Tevatron, with a yearly integrated luminosity
of $2 \times 10^3 \ pb^{-1}$, then the number of events will be about 50,
which seems quite likely to be detected because the signal is almost 
background free.

The enhancement in the cross-section could increase 
significantly the detection feasibility of the Higgs boson
at the LHC. The most promising candidate for the final signature
is again through the decay into a photon pair.
If both $\sigma$ and the B.R. receive a similar
enhancement (of about 5 times), then the number of events 
coming from $pp \to H + X \to \gamma \gamma+X$ may be about 25 
times larger than the SM case, which should be then useful in order to
detect the Higgs boson with the resolution expected for
the invariant mass of photon pairs at the LHC \cite{hixhunt}.

The possibility of detecting a Higgs boson through 
the dominant decay mode ($H \to b\bar{b}$) is unlikely
because of the QCD backgrounds, at least for the gluon
fusion production, nor is the signature coming from
the rare decay $H \to \mu^+ \mu^-$.
Although the branching ratio may be   
enhanced by an order of magnitude, as it was discussed 
in detail in \cite{toscab}, the event rate will not 
be large enough to compete with the background coming 
from the decay of the Z into lepton pairs \cite{hixhunt}.

On the other hand, the operator $O^{6,l}_{\phi Gt}$ gives also a 
significant contribution to the gluon fusion cross-section. 
In case this effect were observed it would signal new physics.
We can also use this result to put a bound on the
coefficient $\eta_{\phi G t}$. 
Using a simple perturbative constraint, 
namely that the effect of the new operator 
should not be larger than the SM result, we obtain the bound 
$\eta_{\phi G t}< 6.8 (\Lambda/TeV)^2$. This in turn
can be transformed into a bound on the 
chromomagnetic dipole moment ($\mu_t$)
for the top quark \cite{topeff}, 

\begin{equation}
\hat{\mu_t}=  \frac{m_t v \eta_{\phi G t} }{8\pi^2 \Lambda^2},
\end{equation}
with $|\hat{\mu_t}| < 2.4 \times 10^{-3}$, $\hat{\mu_t} = m_t\mu_t/g_s$,
and $\mu_t$ is the coefficient associated to the term 
$\bar{t} G^{\mu \nu} \sigma_{\mu \nu} t$, which in turn  comes from
the effective operator (3). Note that our result given in (14) includes
an additional suppressing $v/\Lambda$ factor because the effective
operator (3) is of dimension 6.
This bound is stronger than the one obtained previously, 
$\hat{\mu_t} < 0 (1)$, by the use of $U(1)_{em}$ gauge 
invariant $t\bar t G$ couplings on the 
$t\bar t$ production cross section in hadron-hadron collisions \cite{topeff}.

The previous operators (2) and (3) may also 
modify the cross-section for the production of
the Higgs boson in association with a $t\bar{t}$ pair, 
which is believed to be the most viable reaction 
for detection of a Higgs boson in the 
intermediate-mass region.
A similar analysis lead us to conclude that
the dominant effect will arise from the operator
$O^{6,t}_{t\phi }$, whose effect may be extracted from 
the following ratio:
\begin{eqnarray}
R_{Ht\bar{t}}&=&\frac{\sigma_{eff}(pp \to H+t\bar{t}+X)}
             {\sigma_{SM}(pp \to H +t\bar{t}+X)} \nonumber \\
      &=& 1-\frac{3}{2\sqrt{2}}z_1^2z_2\alpha^{6,t}_{t\phi}.   
\end{eqnarray}
In this case we find that the largest increase in the 
corresponding cross-section may be of the order of 20 percent.
If we consider again 
the most viable signal, namely through the decay of the
Higgs into a photon pair, the increase in the final 
event rate could be about 100 percent, which should be detectable at
the LHC since the SM signal was found already detectable \cite{hixhunt}.
Under these circumstances the signals coming from the SM and the effective
theory will be
disentangled  provided that the event rate is large enough.
Eventhought $R_{H\bar{t}t} \ = \ 2$ seems clearly
distinguishable, the precise confident level depends in general on 
the Higgs mass.

Finally, we have also studied  
the modifications to the mechanism of
associated production of Higgs with a W boson,
due to the new 6-dimensional operators. This reaction 
plays also a very important role for the detection of
a Higgs boson in the intermediate-mass region.
The Feynman graphs for this process
are shown in Fig. 2. Again,
the dots denote the vertices induced by the new operators
in the effective Lagrangian. From the
discussion of the previous section, we can appreciate that the dominant
contribution will come from the lowest dimension tree-level  
induced operators, which are the dimension-6
operators $O^{6,t}_{\phi1,3 }$ that modify the 
interaction $HWW$. The operators that modify the 
$qqW$ vertex are one-loop induced operators 
and can be neglected.

We have evaluated the ratio of the effective cross-sections 
to the SM result. The form of
the operators is such that the parton convolution part is
factored out and we are left with only the ratio of partonic 
cross-sections. Thus our result is
valid for both the FNAL and LHC cross sections,

\begin{eqnarray}
R_{HW}&=&\frac{\sigma_{eff}(pp \to H+W+X)}{\sigma_{SM}(pp \to H +W+X)} \\
      &=& 1+0.5 (2\alpha_{\phi 1}-\alpha_{\phi 3})z^2_1,   
\end{eqnarray}
with $z_1$ defined in (10). 
For this type of operators the modification to the 
cross-section is independent of the Higgs mass.
We find that the best values for the cross-section are only
slightly modified. For instance, with $m_H=100$ GeV 
we get that the ratio lies in the range:
$0.903 < R_{HW} < 1.097$, and thus the chances to detect the Higgs boson
remain as good as in the SM, but does not seems likely to distinguish the
effective theory signal from the SM one.

{\bf 5.- Conclusions.}  We have studied
the possibility to detect heavy physics in
the interactions of the Higgs boson and the top quark at future
colliders using the effective Lagrangian 
approach. It was assumed that the SM lagrangian is modified
by non-renormalizable operators that are invariant under the
full strong and electroweak group. We found that the lowest-order operators
that modify the SM $t\bar{t}g$ strong interaction are
of dimension 6.

The modification to the Higgs
and top interactions may enhance the Higgs
production at hadron colliders, mainly through the mechanisms
of gluon fusion and associated production with a $t\bar{t}$
pair. It is found that the operator  $O^{8,t}_{\phi Gt}$
may lead to an enhancement in the cross-section by a
factor of 5.3, 
which increases the possibilities to detect the Higgs boson
at future colliders. It is found that the most promising
signature comes from the decay into a gamma pair, which may
receive a similar enhancement by the effective operators
and makes plausible its detection at the LHC.

Similarly, the mechanism of associated production of the Higgs
with a $t\bar{t}$ pair is also enhanced, this time only by about 
20 percent. On the other hand, the
associated production of the Higgs with a W boson
is not modified substantially, which means that the possibilities 
to detect this signal are as good as in the SM, or in other
words: {\it{if a light Higgs boson exists, then its detection
through this mode is quite model independent}} .

We have also analyzed the possibility to use the gluon fusion 
mechanism to obtain a bound on the  
chromomagnetic dipole moment of the top. 
Using a simple perturbative constraint, namely that the effect of 
the new operator should not be larger than the SM result,
we obtain the bound 
$\eta_{\phi G t}< 6.8 (\Lambda/TeV)^2$, which leads to 
a stronger bound on $\hat \mu_t$ than obtained previously 
in the literature.

{{\bf Acknowledgement}. We acknowledge financial support
from CONACYT and SNI (M\'exico).}
\newpage



\newpage

\bigskip

{\bf FIGURE CAPTION}

\bigskip

\bigskip

{\bf Fig. 1} Feynman graphs that contribute to the  
loop induced $Hgg$ interaction. 

\bigskip

{\bf Fig. 2} Feynman graphs that contribute to the    
mechanism of H and W associated production. 

\bigskip

\bigskip

{\bf TABLE CAPTION}

\bigskip

\bigskip

{\bf Table 1} Results for the contribution of the effective operators  
to the ratio $R_{Hgg}$. We have taken $m_H=100$ GeV, and $\Lambda \ = \ 1$ 
TeV. The total value is obtained by adding the amplitudes arising from
each operator, and then squaring the result.

\newpage

\begin{center}
\begin{tabular}{||l|l|l||}
\hline
  & \multicolumn{2}{c||}{Ratio} \\
\cline{2-3}
Operators  & $R^{max}_{Hgg}$ & $R^{min}_{Hgg}$  \\
\hline
$O^{6,t}_{t\phi }$   & 1.204  & 0.816 \\  
\hline
$O^{6,l}_{\phi tG }$ & 1.325  & 0.721 \\  
\hline
$O^{6,l}_{\phi G }$  & 1.199  & 0.819 \\  
\hline
$O^{8,t}_{\phi G }$  & 3.873  & 0.01 \\  
\hline
Total                & 5.339  & $7.1 \times 10^{-5}$ \\  
\hline
\end{tabular}
\end{center}

\centerline{Table 1}

\end{document}